\documentclass[review]{elsarticle}

\usepackage{lineno,hyperref}

\usepackage{graphicx}
\usepackage{amssymb}

\usepackage{lineno}
\journal{Journal Name}

\begin{document}

\begin{frontmatter}


\title{Predication of Final Medal Counts in Olympic Games by Monte Carlo Simulations}



\author{Maggie Barker\footnote{mab4758@uncw.edu}, Daniel Guo\footnote{guod@uncw.edu, corresponding author.}, Justin Palmeri\footnote{jtp1895@uncw.edu}, and Ridge Shepherd\footnote{wrs6026@uncw.edu}}

\address{Department of Mathematics \& Statistics \\ University of North Carolina Wilmington, United States}

\begin{abstract}
In the paper, a program strength model was proposed to evaluate the performance of countries across different Olympic events. The model assessed how strong a country’s program was in each event and also factored in the influence of past Olympic performances. The final medal counts from the Paris 2024 Olympic Games were used to validate the model and to determine the optimal set of constants using Monte Carlo simulation. Based on this model, a prediction of the final medal counts for the 2028 Olympic Games is also provided for reference.
\end{abstract}


\end{frontmatter}


\section{Introduction}
The Summer Olympic Games have evolved significantly since their modest beginnings in Ancient Greece, where they first took place in 1895. Back then, only 241 competitors from 14 nations participated. By Paris 2024, the Games had grown into a global spectacle, featuring 1,074 athletes representing 206 nations or regions \cite{OlympicData2024}. Beyond showcasing athletic excellence, the Olympics continue to inspire millions around the world, promoting international unity and often reflecting the political and social dynamics of their time. Moreover, the Games leave a lasting impact on host cities and countries, driving investment in sports, culture, and infrastructure development.

Every four years, the world watches the Summer Olympic Games with excitement and anticipation, cheering on their nations and favorite athletes across a wide range of sporting events. Winning an Olympic medal is widely regarded as the pinnacle of athletic achievement, and countries invest substantial time and resources into training athletes to compete at this elite level. For many athletes, simply qualifying for an Olympic event is a lifelong dream.

In addition to athletic preparation, host countries invest heavily in building venues and providing accommodations for fans, athletes, coaches, journalists, performers, and other participants. While some aspects of the Games' outcomes may be influenced by chance or unpredictable events, sports forecasters are increasingly able to predict results with a degree of accuracy \cite{Condon1999, Ogwang2021, Radicchi2012}. These predictions serve as important tools for nations seeking to evaluate their performance after the competition.

Over the years, researchers have developed various mathematical models to forecast Olympic medal counts, employing a wide range of methodologies. These models often incorporate diverse factors, such as historical performance data, population sizes, and national resources. For example, population-adjusted approaches \cite{Duncan2024} account for the relative size of a country's population, positing that larger nations may have more athletes and, therefore, a greater potential to win medals. Similarly, machine learning techniques \cite{Schlembach2022} have gained traction in recent years, utilizing vast datasets to uncover patterns and make predictions with increasing accuracy.

Despite the advancements in forecasting models, there remains room for improvement in the precision and reliability of medal predictions. Many existing models still struggle to fully capture the complexity of the factors that contribute to Olympic success, such as the strength and depth of a country's sports programs, changes in training methodologies, and the unpredictable nature of the competition itself.

In this paper, we propose a novel mathematical model that integrates both historical performance data and program strength, two critical yet often under-emphasized factors. By examining trends in past Olympic performances and evaluating the investments made by countries in their sports infrastructure, we aim to provide a more nuanced and comprehensive prediction model. This model not only takes into account the overall medal counts of previous Games but also factors in the improvement and evolution of a nation's sporting programs over time, thus offering a more dynamic approach to forecasting Olympic success.

By refining and expanding upon existing models, our approach aims to offer more accurate and actionable predictions for policymakers, athletes, and sports analysts. These predictions could help countries evaluate their preparedness for upcoming Olympics, guide strategic investments in sports development, and provide valuable insights into the evolving dynamics of global athletic competition.

In this paper, we use the final medal standings from the Paris 2024 Olympics to train our model and determine key parameters. This data serves as the foundation for our predictive framework, allowing us to fine-tune the model and ensure its accuracy in forecasting future Olympic outcomes. We apply the program strength model, which takes into account the historical performance data, the investment in sports infrastructure, and the depth of athletic talent within each country or region. We incorporate a Monte Carlo simulation, a statistical technique that accounts for uncertainty and variability in the forecasting process.

By combining these methods, we predict the final medal counts for the 2028 Olympic Games. The prediction table, which lists the expected medal standings for each country or region, is included for reference. These predictions offer valuable insights into the likely distribution of medals and can inform decision-making for athletes, coaches, and policymakers in their preparations for the upcoming Games.

\section{The Program Strength Model}
\label{S:4}
We use historical data \cite{OlympicData2024} to inform our predictions and propose a new model to estimate the strength of each country’s performance in individual Olympic events for future Games. The model takes into account the number of athletes who earned gold, silver, and bronze medals, as well as those who participated without winning a medal. Each outcome is assigned a corresponding point value. These points are then weighted by a diminishing time factor, giving more recent results greater influence. The weighted points are summed to produce an overall strength score for each country in each event.

Based on these strength scores, we analyze the differences between the top two or three countries in each event. If the gap between them is small, we consider the possibility of a country winning multiple medals—referred to as “double medals”—in that event, especially for the top-ranked or second-ranked countries. 

In this model, only historical data \cite{OlympicData2024}, such as medal distributions and the number of participants, is used to inform predictions. Each data point contributes to the projected medal outcomes based on assigned weights, which reflect its relative importance. Naturally, if more detailed data—such as athletes’ ages or career stages—were available, the model’s accuracy could be further improved.

Since the Olympic Games are held every four years, recent performances have a stronger influence on future outcomes, while older results are given less weight. This reflects the evolving nature of athlete development and national sports programs. While our current model relies solely on Olympic data, it could be enhanced by incorporating results from recent international competitions, which might provide additional predictive power.

Ultimately, the final medal counts are determined by performance in individual events. Our approach calculates a strength score for each participating country in every event, and medals are then assigned to the top-ranked countries based on these scores

Let $x$ be a country and $P_{ix}$ be the scored points for the event for the country $x$. Then define the score as
\begin{equation}
P_{ix} = \sum^n_{j=1} \frac{0.8}{2^j} (a G_{ijx}+b S_{ijx}+c B_{ijx}+d Y_{ijx})
\label{model1}
\end{equation}
where
\begin{itemize}
    \item $i$ is the any event of Olympic Games.
    \item $n$ is the total Olympic Games to be considered.
    \item $G_{ijx}, S_{ijx}, B_{ijx}$ are medals at $jth$ Olympic Game for the country $x$ at the $ith$ event. They are either $0$ or $1$.
    \item $Y_{ijx}$ is the total players at $jth$ Olympic Games for the country $x$ at the $ith$ event.
    \item $a, b, c, d$ are constants to be decided later.
\end{itemize}

Theoretically,  we could consider all previous Olympic Games. However, the test results showed less impact for the data twenty-four year ago. 

Now define the medal count $M_{ix}$ of the country $x$ for the event $i$ as follows, for simplicity, $M_{ix}$ is $0$ if $P_{ix}$ is not on the top three score list. More accurately, we consider the possibility of multiple medals. Let $P_{ix}$ be the top score for the event $i$,
\begin{itemize}
    \item If $P_{ix}$ is larger than the double of the sum of the next two top scores, then $M_{ix}=3$.
    \item If $P_{ix}$ is larger than the sum of the next two top scores or double the the next top score, then $M_{ix}=2$.
    \item otherwise, $M_{ix}=1$.
\end{itemize}

The total predicted final medal count for the country $x$ is defined by
$$ PM_x = \sum_i M_{ix}.$$

We want to find the constants $a, b, c, d$ such that the error $E$ is minimized as
$$E = \sum_x (PM_x-AM_x)^2 $$
where $AM_x$ is total actual medal count for the country $x$.

\section{Monte Carlo Simulations}
In order to decide the constants $a, b, c, d$, we used the actual final medal counts of Olympic Games Paris 2024 as the reference data ($AM$). It is possible to use the final medal counts of any year as the reference data. However, the most recent data as the reference data will generate better prediction results for the future Olympic Games. 

Those constants could be any positive numbers theoretically. For the simplicity, we restricted them in the box as $0 \le a, b, c, d \le 10$. Then using Monte Carlo simulations to minimize $E$. The best five choices are listed in the table 1 over 5000 samples.

\begin{table}[ht]
\caption{Possible best five choices of constants for year 1980 to year 2020 } \vspace{0.1in}
\centering
\begin{tabular}{|c|c|c|c|c|}
\hline
a & b & c & d & E \\ \hline
5.3371	& 5.2041	& 8.0534	& 0.1926	& 297 \\\hline
5.3938	& 5.4786	& 8.4638	& 0.4358	& 297 \\\hline
5.4667	& 5.5657	& 8.5813	& 0.3120	& 298 \\\hline
4.7707	& 4.8672	& 7.2926	& 0.3672	& 298 \\\hline
4.1996	& 4.2017	& 6.5660	& 0.1842	& 302 \\\hline
%
\end{tabular}
\end{table}

\begin{table}[h!]
\centering
\caption{Predicted vs Actual Final Medal Counts with Errors for Olympic Games Paris 2024}\vspace{0.1in}
\begin{tabular}{|l|c|c|c|}
\hline
\textbf{Team} & \textbf{Predicted} & \textbf{Actual} & \textbf{Error}\\
\hline
United States & 127 & 126 & 1 \\
China & 89 & 91 & 2 \\
Great Britain & 69 & 65 & 4 \\
France & 63 & 64 & 1 \\
Australia & 57 & 53 & 4 \\
Japan & 50 & 45 & 5 \\
Italy & 43 & 40 & 3 \\
Netherlands & 38 & 34 & 4 \\
Germany & 35 & 35 & 0 \\
Canada & 30 & 27 & 3 \\
South Korea & 27 & 32 & 5 \\
Spain & 22 & 18 & 4 \\
New Zealand & 22 & 20 & 2 \\
Brazil & 21 & 20 & 1 \\
Hungary & 19 & 19 & 0 \\
\hline
\end{tabular}
\end{table}

Using the dataset \cite{Schlembach2022}, we found the best choice of $a, b, c, d$ as $5.337073971$, $5.204062091$,  $8.053396832$, and $0.192604723$. With these values, it takes the bronze medals into consideration the most with a value of $8.053396832$, while silver is taken into account almost at the same value for gold and then the overall athletes are only slightly taken into consideration with a value of $0.192604723$.

We considered all events in Olympic Games Paris 2024 including five special sports added by the host country. The comparison of the final medal counts of Olympic Games Paris 2024 is listed in the table 2 for top 15 countries. The largest difference in final medal counts is 5.

As the comparison, we tested two sets of data. One is from year 1980 to year 2020, and the other one is from year 2000 to year 2020. After $5000$ sampling, the top ten results are almost same with the first set of data is slightly better. The top 5 chooses for the data set from 2000 to year 2020 are listed in the table 3.
\begin{table}[ht]
\caption{Possible best five choices of constants for year 2000 to year 2020} \vspace{0.1in}
\centering
\begin{tabular}{|c|c|c|c|c|}
\hline
a & b & c & d & E \\ \hline
6.5804 & 6.8997 & 8.6227 & 0.2729 & 300 \\\hline
5.4190 & 4.7867 & 7.5191 & 0.2928 & 301 \\\hline
4.9823 & 4.8924 & 7.7206 & 0.3694 & 302 \\\hline
5.5178 & 5.4439 & 8.5962 & 0.4043 &	302 \\\hline
5.8051 & 5.6236 & 8.7725 & 0.3367 &	302 \\\hline
\end{tabular}
\end{table}

\section{Prediction of the final metal counts of Olympic Games 2028}
Apply the best choice of the constants from the table 1, we have the model as
$$
P_{ix} = \sum^n_{j=1} \frac{0.8}{2^j} (a G_{ijx}+ b S_{ijx}+ c B_{ijx}+ d Y_{ijx})
$$
with $a=5.337073971$, $b=5.204062091$,  $c=8.053396832$, and $d=0.192604723$. 
where all indexes are defined as in the equation (\ref{model1}). 

For the prediction of the final medal counts in the 2028 Olympic Games, we consider all countries and regions that participated in the 2024 Olympics. Additionally, we use data from the years 1980 to 2024 as part of our analysis. The top 10 medal-winning countries or regions are listed here for reference. The full list of predicted medal standings can be found in the appendix. It is important to note that the final medal counts exclude medals from the five newly added sports, as there is insufficient historical data available for these sports. Consequently, the actual final medal counts are expected to be higher than those shown in the table.

For computational efficiency, we can use data from the years 2000 to 2024 to make our predictions for the 2028 Olympic Games medal counts. The results are little different, but are very close.

\begin{table}[ht]
\centering
\caption{Predictions of Final Medal Counts in 2028 Olympic Games} \vspace{0.1in}
\begin{tabular}{|l|c|c|c|c|}
\hline
\textbf{Country} & \textbf{Gold} & \textbf{Silver} & \textbf{Bronze} & \textbf{Total} \\
\hline
United States & 62 & 31 & 31 & 124 \\
China & 45 & 20 & 23 & 88 \\
Great Britain & 25 & 31 & 13 & 69 \\
France & 15 & 18 & 29 & 62 \\
Australia & 18 & 20 & 19 & 57 \\
Japan & 12 & 21 & 13 & 46 \\
Italy & 13 & 16 & 12 & 41 \\
Netherlands & 13 & 12 & 13 & 38 \\
Germany & 8 & 16 & 11 & 35 \\
Canada & 7 & 8 & 14 & 29 \\
\hline
\end{tabular}
\end{table}

\section{Conclusions}
The program strength model offers a reliable, data-driven approach to predicting Olympic outcomes for 2028 based on historical performance. By focusing on measurable factors and avoiding uncertain variables, the model provides accurate insights that can help anticipate the potential performances of nations in the next Olympic Games.

By using data from the 2024 Olympics, the model leverages reliable, real-world performance outcomes that are measurable and verifiable.
Historical data serves as a solid foundation because it reflects consistent patterns and trends in how countries perform across events over time.

The decision to exclude uncertain and unpredictable factors (such as political changes, new sports, or emerging trends) keeps the model simple, yet effective. By avoiding speculative elements, the model stays grounded in what is known, which increases its accuracy and ensures it can provide reliable predictions.

The predictive power of the model lies in its ability to give a sense of how countries might perform in future Olympics, helping stakeholders—such as athletes, coaches, national sports federations, and even fans—prepare for the next Games. 
The program strength model can help identify countries that may be strong contenders, as well as those that may see improvements or declines in their performances.

\section*{Appendix: Complete final medal counts for Olympic Games 2028}

\begin{table}[ht]
\centering
\caption{Predicted Medal Counts by Country (1-25)}
\vspace{0.1in}
\begin{tabular}{|l|c|c|c|c|}
\hline
\textbf{Country} & \textbf{Gold} & \textbf{Silver} & \textbf{Bronze} & \textbf{Total} \\
\hline
USA & 62 &	31 &	31 &	124 \\
China & 45 &	20 &	23	& 88 \\
Great Britain & 25 &	31 &	13 &	69 \\
France & 15 &	18 &	29 &	62 \\
Australia & 18 &	20 &	19 &	57 \\
Japan & 12 &	21 &	13 &	46 \\
Italy & 13 & 16 & 12 & 41 \\
Netherlands & 13 & 12 & 13 & 38 \\
Germany & 8 & 16 & 11 & 35 \\
Canada & 7 & 8 & 14 & 29 \\
South Korea & 11 & 8 & 8 & 27 \\
New Zealand & 6 & 5 & 11 & 22 \\
Brazil & 5 & 10 & 6 & 21 \\
Spain & 12 & 4 & 5 & 21 \\
Hungary & 8 & 4 & 6 & 18 \\
Denmark & 2 & 7 & 3 & 12 \\
Poland & 4 & 4 & 4 & 12 \\
Iran & 1 & 6 & 4 & 11 \\
Kenya & 7 & 2 & 2 & 11 \\
Cuba & 6 & 2 & 2 & 10 \\
Sweden & 1 & 1 & 8 & 10 \\
Ukraine & 2 & 3 & 5 & 10 \\
Uzbekistan & 1 & 4 & 5 & 10 \\
Czechia & 2 & 3 & 4 & 9 \\
Jamaica & 2 & 4 & 3 & 9 \\
\hline
\end{tabular}
\end{table}

\begin{table}[ht]
\centering
\caption{Predicted Medal Counts by Country (26-50)}
\vspace{0.1in}
\begin{tabular}{|l|c|c|c|c|}
\hline
\textbf{Country} & \textbf{Gold} & \textbf{Silver} & \textbf{Bronze} & \textbf{Total} \\
\hline
Switzerland & 3 & 6 & 0 & 9 \\
Croatia & 2 & 2 & 4 & 8 \\
Romania & 0 & 2 & 6 & 8 \\
Taiwan & 2 & 2 & 4 & 8 \\
Austria & 4 & 2 & 1 & 7 \\
Belgium & 3 & 3 & 1 & 7 \\
Greece & 2 & 3 & 2 & 7 \\
India & 4 & 2 & 1 & 7 \\
Kazakhstan & 1 & 2 & 3 & 6 \\
Norway & 2 & 0 & 4 & 6 \\
North Korea & 0 & 3 & 3 & 6 \\
South Africa & 2 & 3 & 1 & 6 \\
Bulgaria & 1 & 4 & 0 & 5 \\
Colombia & 1 & 1 & 3 & 5 \\
Dominican Republic & 0 & 2 & 3 & 5 \\
Ecuador & 2 & 1 & 2 & 5 \\
Ethiopia & 1 & 2 & 2 & 5 \\
Georgia & 3 & 0 & 2 & 5 \\
Israel & 1 & 2 & 2 & 5 \\
Mexico & 1 & 2 & 2 & 5 \\
Slovenia & 0 & 3 & 2 & 5 \\
Serbia & 1 & 2 & 2 & 5 \\
Thailand & 2 & 2 & 1 & 5 \\
Turkey & 1 & 2 & 2 & 5 \\
Armenia & 1 & 1 & 2 & 4 \\
\hline
\end{tabular}
\end{table}

\begin{table}[t]
\centering
\caption{Predicted Medal Counts by Country (51-75)}
\vspace{0.1in}
\begin{tabular}{|l|c|c|c|c|}
\hline
\textbf{Country} & \textbf{Gold} & \textbf{Silver} & \textbf{Bronze} & \textbf{Total} \\
\hline
Azerbaijan & 1 & 0 & 3 & 4 \\
Ireland & 1 & 2 & 1 & 4 \\
Kyrgyzstan & 2 & 1 & 1 & 4 \\
Lithuania & 2 & 1 & 1 & 4 \\
Portugal & 1 & 2 & 1 & 4 \\
Argentina & 1 & 0 & 2 & 3 \\
Grenada & 1 & 1 & 1 & 3 \\
Hong Kong & 1 & 1 & 1 & 3 \\
Indonesia & 1 & 2 & 0 & 3 \\
Philippines & 1 & 1 & 1 & 3 \\
Tajikistan & 0 & 1 & 2 & 3 \\
Antigua \& Barbuda & 0 & 2 & 0 & 2 \\
Algeria & 0 & 1 & 1 & 2 \\
Belarus & 0 & 0 & 2 & 2 \\
Botswana & 0 & 1 & 1 & 2 \\
Bahrain & 1 & 0 & 1 & 2 \\
Ivory Coast & 0 & 1 & 1 & 2 \\
Egypt & 1 & 1 & 0 & 2 \\
Kosovo & 0 & 0 & 2 & 2 \\
Saint Lucia & 0 & 0 & 2 & 2 \\
Morocco & 0 & 2 & 0 & 2 \\
Malaysia & 1 & 1 & 0 & 2 \\
Moldova & 1 & 1 & 0 & 2 \\
Puerto Rico & 0 & 1 & 1 & 2 \\
Qatar & 1 & 1 & 0 & 2 \\
\hline
\end{tabular}
\end{table}

\begin{table}[t]
\centering
\caption{Predicted Medal Counts by Country (76-96)}
\vspace{0.1in}
\begin{tabular}{|l|c|c|c|c|}
\hline
\textbf{Country} & \textbf{Gold} & \textbf{Silver} & \textbf{Bronze} & \textbf{Total} \\
\hline
Russia & 0 & 0 & 2 & 2 \\
Tunisia & 0 & 1 & 1 & 2 \\
Uganda & 0 & 2 & 0 & 2 \\
Albania & 0 & 0 & 1 & 1 \\
Chile & 0 & 0 & 1 & 1 \\
Cape Verde & 0 & 0 & 1 & 1 \\
Cyprus & 0 & 0 & 1 & 1 \\
Dominica & 0 & 0 & 1 & 1 \\
Eritrea & 0 & 0 & 1 & 1 \\
Fiji & 0 & 1 & 0 & 1 \\
Guatemala & 0 & 1 & 0 & 1 \\
Kuwait & 0 & 0 & 1 & 1 \\
Latvia & 0 & 1 & 0 & 1 \\
Mongolia & 0 & 1 & 0 & 1 \\
Nigeria & 0 & 0 & 1 & 1 \\
Peru & 0 & 1 & 0 & 1 \\
Singapore & 1 & 0 & 0 & 1 \\
San Marino & 0 & 0 & 1 & 1 \\
Slovakia & 1 & 0 & 0 & 1 \\
Venezuela & 0 & 0 & 1 & 1 \\
Zambia & 0 & 1 & 0 & 1 \\
\hline
\end{tabular}
\end{table}


\end{document}